\documentclass[prl,twocolumn,superscriptaddress,floatfix,showpacs,showkeys,preprintnumbers,amssymb
,amsmath]{revtex4}
\usepackage{graphicx}
\usepackage{dcolumn}
\usepackage{bm}
\usepackage[latin1]{inputenc}
\usepackage[mathscr]{eucal}
\usepackage{epsfig}
\usepackage{rotating}

\begin{document}
\preprint{ }

\title{Observation of transition from escape dynamics to underdamped phase diffusion in a Josephson junction}

\author{J.M. Kivioja}
\email{jkivioja@boojum.hut.fi} \affiliation{Low Temperature
Laboratory, Helsinki University of Technology, POB 3500, FIN-02015
HUT, Finland}

\author{T.E. Nieminen}
\affiliation{Low Temperature Laboratory, Helsinki University of
Technology, POB 3500, FIN-02015 HUT, Finland}

\author{ J. Claudon}
\affiliation{Centre de Recherches sur les Tr\`es Basses
Temp\'eratures, laboratoire associ\'e \`a l'Universit\'e Joseph
Fourier, C.N.R.S., BP 166, 38042 Grenoble-cedex 9, France}

\author{O. Buisson}
\affiliation{Centre de Recherches sur les Tr\`es Basses
Temp\'eratures, laboratoire associ\'e \`a l'Universit\'e Joseph
Fourier, C.N.R.S., BP 166, 38042 Grenoble-cedex 9, France}

\author{ F.W.J. Hekking}
\affiliation{ Laboratoire de Physique et Mod\'elisation des
Milieux Condens\'es, C.N.R.S. and Universit\'e Joseph Fourier,
B.P. 166, 38042 Grenoble Cedex 9, France}

\author{J.P. Pekola}
\affiliation{Low Temperature Laboratory, Helsinki University of
Technology, POB 3500, FIN-02015 HUT, Finland}

\date{\today}

\begin{abstract}

We have investigated the dynamics of underdamped Josephson
junctions. In addition to the usual crossover between macroscopic quantum tunnelling
and thermally activated (TA) behaviour we observe in our samples
with relatively small Josephson coupling $E_J$, for the first time, the transition 
from TA behaviour to underdamped phase diffusion.
Above the crossover temperature 
the threshold for switching into the finite voltage state becomes extremely sharp. 
We propose a ($T,E_J$) phase-diagram with various regimes and show that for a proper
description of it dissipation and level
quantization in a metastable well are crucial.

\end{abstract}

\pacs{74.50.+r, 85.35.-p,85.25.Dq} \keywords{Josephson junction,
phase diffusion, energy level quantization}

\maketitle

A hysteretic Josephson junction (JJ) switching from its metastable zero-voltage state
into a stable state with a voltage of the order of twice the superconducting gap
$\Delta$ has recently been used as a read-out device for superconducting quantum bit
systems in many experiments \cite{Vion02}. Switching measurements also enable one to
perform conventional large bandwidth current measurements, and recently there have
been proposals to use hysteretic JJs as ammeters for studying phenomena like
non-Gaussian noise \cite{Tobiska03}. Often it may be advantageous to reduce the
critical current $I_c$ of the detecting junction in order to increase the measurement
sensivity. Yet the physics governing switching phenomena of small $I_c$ junctions
ultimately differs from those with larger $I_c$, and this sets a limit on how far one
can reduce $I_c$ still maintaining the useful features of the detector \cite{Vion,
HTC}.

\begin{figure}[h]
\begin{center}
\resizebox{.4\textwidth}{!}{\includegraphics{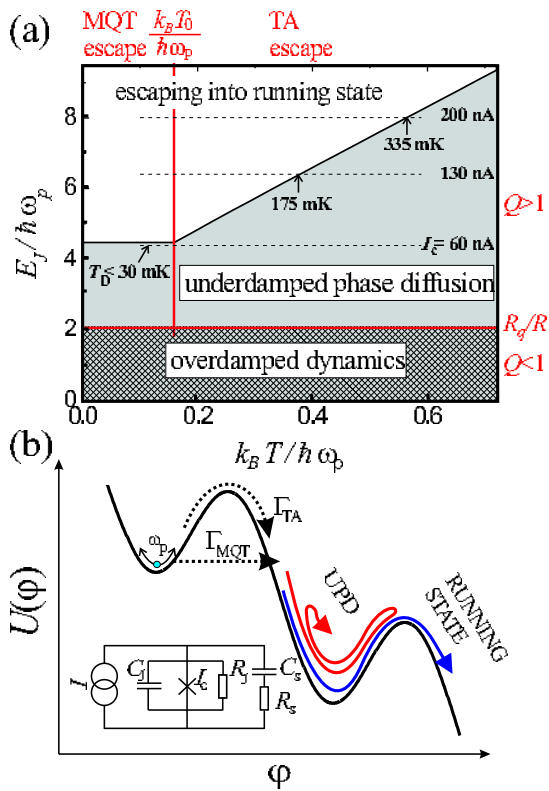}}
\caption{\label{fig:phasediagram} (a) The various operation
regimes of a Josephson junction with low $E_J$. For details, see
text. (b) The cosine potential with dynamics {\em inside} the
upper well and schematic dynamics \emph{after} leaving the upper
well. Inset: equivalent circuit of the junction with frequency
dependent dissipation.}
\end{center}
\end{figure}

The behavior of a JJ is determined by various parameters. Within
the RCSJ (Resistively and Capacitively Shunted Junction) model
these are the Josephson energy $E_J = \hbar I_c/2e$, the charging
energy $E_C = e^2 /2C_J$ ($C_J$ is the junction capacitance), and
a shunt resistance $R$ responsible for dissipation. The dynamics
of the junction are then characterized by the plasma frequency $
\omega_p = \sqrt{8E_J E_C}/\hbar$ and the quality factor $Q= \omega
_p RC$. As we will detail below, the behavior of the junction at a
given temperature $T$ can be classified according to the phase
diagram of Fig.~\ref{fig:phasediagram}a. The overdamped case $Q<1$
was studied in detail by Vion {\em et al.}~\cite{Vion}, who
uncovered the existence of a phase diffusion regime at finite $T$
with the appearance of a small voltage, prior to switching to a
voltage of the order of $2 \Delta$. As far as the underdamped case
$Q > 1$ is concerned, most experiments were done on large
hysteretic junctions with relatively high $I_c$. Depending on
temperature such junctions escape out of the metastable
zero-voltage state either via macroscopic quantum tunnelling (MQT)
or thermal activation (TA) processes, thereby switching directly
to the finite-voltage state. In this Letter we focus on the regime
$Q>1$ with relatively small $I_c$. We show, theoretically and
experimentally, that a regime exists where escape does not lead to
a finite voltage state but rather to underdamped phase diffusion
(UPD, the shaded region in Fig.~\ref{fig:phasediagram}).

According to the RCSJ model, the dynamics of a JJ biased with a
current $I$ is that of a phase particle (phase $\varphi$) with
mass $m=\hbar ^2/8 E_C$, moving in a tilted cosine potential $
U(\varphi)=-E_J \;( {\rm cos}\varphi+I/I_c\;\varphi)$ under a
viscous force $ (\frac{\hbar}{2e})^2 \; 1/R \; \frac{d
\varphi}{dt}$.
The bias current renormalizes the plasma frequency, such that
$\omega _p=\sqrt{\frac{ d^2 U/dq^2}
{m}}=\frac{1}{\hbar}\sqrt{8E_JE_C}\;q_0^{1/2}$ with
$q_0=\sqrt{2(1-I/I_c)}$. The cosine potential has wells where the
phase particle can be localized; the phase then has constant
average value, and the average voltage across the junction is
zero. For non-zero $I$ the quantum levels in the potential well are metastable and the
particle can escape from a given well either via TA over, or MQT
through the barrier. For large junctions
$\Gamma _{TA}= \frac{\omega _p}{2 \pi}\; e^{-\frac{\Delta U}{k_B
T}}$ for the TA escape rate and $ \Gamma _{MQT} =  12 \sqrt{6 \pi}
\frac{\omega _p}{2 \pi} \sqrt{\Delta U/\hbar \omega _p}
e^{-\frac{36}{5}\frac{\Delta U}{\hbar \omega _p}}$ for the MQT
rate \cite{Weiss}. In the cubic approximation the barrier height
is $\Delta U=\frac{2}{3}\;E_J\;q_0^3$. Below the cross-over
temperature $T_0 =\hbar \omega _p / 2 \pi k_B$ the dominant escape
mechanism is MQT.
The total escape rate is $\Gamma (I)\simeq\Gamma _{MQT} (I) +
\Gamma _{TA}(I) $ and the total escape probability in the time
interval $0\leq t\leq \tau$ can be written as $P(I)=1-e^{-\int
_{0}^{\tau} \Gamma (I) dt}$. If dissipation is weak, upon escape
from the well the particle moves down the potential and phase is
running freely, hence the voltage reaches a finite value (about
$2\Delta^{Al} _{BCS}$$\approx\rm{360\;\mu V}$). However, if
dissipation is strong enough, there is a finite probability that,
upon escape from the well, the particle is relocalized in the next
well instead of running down the potential: the phase then has
$2\pi$-slips and diffusively moves from one metastable well to
another. In this UPD regime the average voltage across the
junction is much smaller than $2 \Delta$~\cite{Vion}.

A misconception still persists that phase diffusion is absent in
a hysteretic junction, although it was pointed out
over a decade ago that it can occur due to the dependence of
dissipation on frequency $\omega$ \cite{Martinis89}. Our
experiment corresponds to the simplified equivalent circuit with
frequency dependent dissipation as presented in the inset of Fig.
\ref{fig:phasediagram}(b). After switching to the running state,
the dominant part of dissipation comes from small $\omega$,
governed by $R(\omega \sim 0)$, typically given by the large
junction subgap resistance, of the order of $\rm{1\;M\Omega}$. In
the phase diffusion regime the phase mainly oscillates in a well
at the plasma frequency and thus the dissipation is characterized
by $R(\omega _p)$, which is much smaller, typically
of the order of vacuum impedance $Z_0\approx\rm{377\;\Omega}$, since $C_s$ acts as a
short. In the following we will consider junctions that are
underdamped even at $\omega _p$, in contrast to~\cite{Vion}.


The dissipated energy between neighboring potential maxima can be approximated by
$E_D\approx 8 E_J/Q$ and if the phase particle has energy less than $E_D$ above the
\emph{next} barrier top, it simply diffuses to the next well. The maximum possible
dissipated power due to phase diffusion can be written as
$\frac {1}{2\pi} \frac{2eV}{\hbar}E_D$, where
$V$ is the average voltage across the junction. By equating
this with the applied bias power $I_m V$, we find the maximum
possible phase diffusion current
\begin{equation}
I_m =\frac{4}{\pi Q}I_c,
    \label{eq:Im}
\end{equation}
which is identical in form to the well known retrapping current
formula, but now the value of $Q$ is that at plasma frequency
$\omega _p$. For $I <I_m$, there is non-zero probability that phase
relocalizes after escape.
With decreasing $I_c$, dissipation starts to play a more
significant role. First, $\omega_p$ and hence the quality factor
$Q$ are decreasing with decreasing $E_J$.
On the other hand the escape rate is significant in the range of
currents where the barrier height is comparable to the thermal
energy. For large junctions, $E_J \gg k_B T$, this is the case
when $I$ is just slightly below $I_c$, but for small
junctions  this occurs even \emph{without} tilting  the cosine
potential ($I/I_c$=0). At small values of $I_c$ the dissipation
\emph{after} escape is thus also important, because the successive
barrier tops are close in energy. The grey area in Fig.
\ref{fig:phasediagram}(a) presents the UPD regime, where escape
does not necessarily lead to the transition into a running state.
The minimum $E_J^D$, where such a transition after escape is still
certain at temperature $T_D > T_0$ with current pulses of length
$\tau$,
\begin{equation}
E_J^D\approx\frac{3}{2}k_B T_D  (1-I_m/I_c)^{-3/2}\rm{ln}(\omega
_p \tau/2\pi ), \label{eq:Ejb}
\end{equation}
determines the line separating TA and UPD regions in Fig.
\ref{fig:phasediagram}(a).
For $T < T_0$, the escape rate equals the TA rate at $T_0$, thus
$E_J^D$ is independent of $T$, and given by Eq.~(\ref{eq:Ejb})
with $T_0$ replacing $T_D$. The UPD region in
Fig.~\ref{fig:phasediagram}(a) corresponds to
$R_s\rm{=500\;\Omega}$, $C_J\rm{=100\;fF}$ and $\tau=\rm{100\;\mu
s}$.

\begin{figure}[!hb]
\begin{center}
\resizebox{.35\textwidth}{!}{\includegraphics{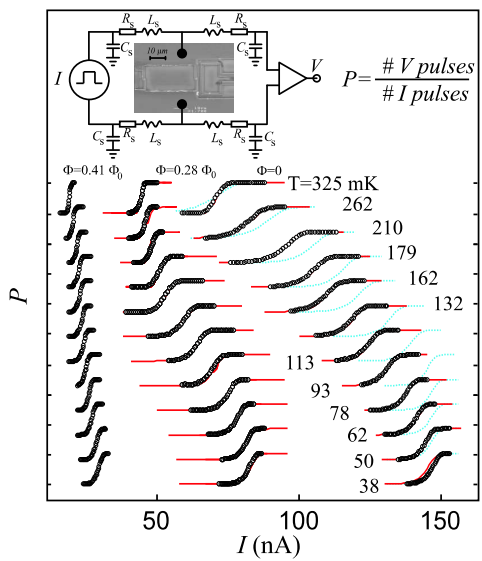}}
\caption{\label{fig:histot} Cumulative histograms of the dc-SQUID
at different temperatures and at different magnetic fields.
Curves are shifted for clarity and the vertical spacing between ticks
corresponds to unity escape probability. Solid lines are from
simulations described in the text; dotted blue lines show the results of basic
MQT and TA models. Inset: scanning electron micrograph of the
measured dc-SQUID and the experimental circuit. }
\end{center}
\end{figure}

We present experimental data of two samples, a dc-SQUID and a single JJ. 
The dc-SQUID consists of two wide superconducting planes connected by two
short superconducting lines with tunnel junctions in the middle forming the dc-SQUID
loop of area $\rm{20\times 39 \;(\mu m)^2}$ (see the inset in Fig. \ref{fig:histot}).
The loop inductance was measured to be around $\rm{100\;pH}$,
small as compared  to the calculated Josephson inductance
($L=\frac{\Phi _0}{2 \pi I_c}=\rm{400\;pH}$  per junction). The
dc-SQUID thus behaves almost like a single JJ,
whose $I_c$ can be tuned. The loop inductances were estimated from
the measured resonant voltage determined by $C_J$
and loop inductance \cite{Zant94}. The other measured
sample was a single junction between long inductive
biasing lines.
The normal state resistances of the dc-SQUID and the single JJ were
$\rm{1.3\;k\Omega}$ and $\rm{0.41\;k\Omega}$ yielding for $I_c$ $\rm{199\;nA}$ and
$\rm{630\;nA}$, respectively. The geometrical capacitances of the samples were
$\rm{100\;fF}$ and $\rm{130\;fF}$. The measured samples were fabricated using standard
electron beam lithography and aluminum metallization in a UHV evaporator.
Both measured samples had strongly hysteretic \emph{IV}-characteristics with
retrapping currents well below $\rm{1\;nA}$.

The experimental setup is presented in the inset in Fig.
\ref{fig:histot}. Switching probabilities have been measured by
applying a set of trapezoidal current pulses through the sample
and by measuring the number of resulting voltage pulses.
At the sample stage we used low pass $RC$-filters (surface mount components near the
sample). In the  measurements on a single junction we used surface mount capacitors
($C_s=\rm{680\;pF}$), but in the dc-SQUID measurements we had $\pi$-filters in series
with resistors , with $C_s\sim\rm{5\;nF}$ capacitance  to ground. The resistors were
$R_s=\rm{500\;\Omega}$ and $\rm{681\;\Omega}$ in the measurement on a dc-SQUID and on
a single junction, respectively. Bonding wires with inductance of order nH connect the
sample to filters.

\begin{figure}[hb]
\begin{center}
\resizebox{.35\textwidth}{!}{\includegraphics{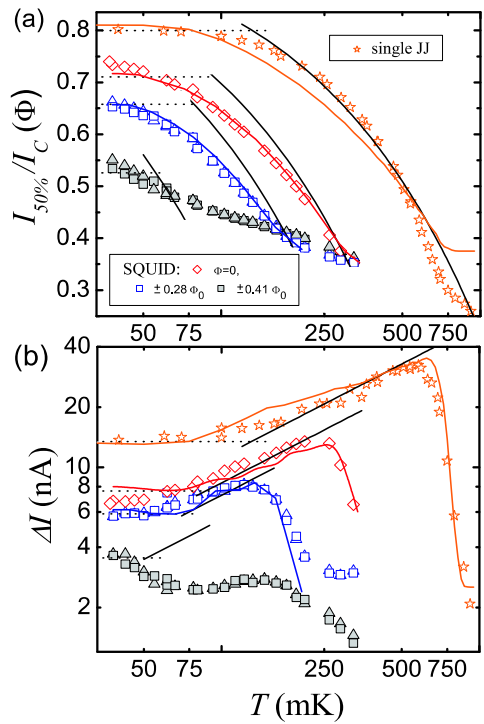}}
\caption{\label{fig:I50&dI} (a) The positions ($I_{50\%}$) and
(b) the widths ($\Delta I$) of the histograms.
Black solid and dotted lines are results (with known junction
parameters) of TA and MQT model, respectively, ignoring
dissipation. Blue, red and orange solid lines are the parameters
of the simulated histograms based on the LO model discussed in the
text. It assumes escape and possible relocalization events to be
rare; thus it is not valid for data at $\Phi=\pm 0.41 \Phi _0$,
where the rate of phase diffusion events approaches the relaxation
in the well.
}
\end{center}
\end{figure}

In Fig.~\ref{fig:histot} we present the measured cumulative
switching histograms (open circles) of the dc-SQUID at different
fluxes $\Phi$ and temperatures with $\tau=200\;\mu s$. At the
lowest temperatures, the histograms can be well fitted by the MQT
model, giving $I_c = \rm{200\;nA}$, $\rm{128\;nA}$ and
$\rm{55\;nA}$, for $\Phi/\Phi_0 = 0$, 0.28 and 0.41 respectively.
For $\Phi = 0$ we also present the simulated histograms based on
the basic TA and MQT models without dissipation (dotted blue lines). In
Fig.~\ref{fig:I50&dI} we show the measured histogram position
$I_{50\%}$ [$P(I_{50\%})\equiv 0.5$] and the width $\Delta I$
($\equiv I_{90\%}-I_{10\%}$) for both samples. The dc-SQUID
measurements were done both at negative and positive values of
flux in order to ensure that the external flux had not changed.
The position in Fig.~\ref{fig:I50&dI}(a) is normalized to the
corresponding value of $I_c$ at zero temperature. We also present
the width and position results of the basic TA and MQT model
simulations. At low $T$ all the measured data are consistent with
MQT results. On increasing $T$ the parameters are constant up to the estimated
cross-over temperature $T_0$. For $T > T_0$ the width is
increasing and the position is moving down as TA model predicts.
The qualitative agreement is good for most of the results up to
the temperature $T_D$. At $T_D$, $\Delta I$ starts to decrease
abruptly. Moreover, the position $I_{50 \%}$ saturates at the same
value $\simeq 0.35 I_c$. For the single JJ, the saturation occurs at
$\simeq 0.3 I_c$. The dc-SQUID data measured at $\Phi=\pm 0.41 \Phi
_0$ are not following the standard theory even at low
temperatures, since $T_D <\rm{30\;mK}$ in this case. If we assume
a realistic shunt impedance $R(\omega _p) \simeq\rm{500\;\Omega}$
(the value of the surface mount resistors), we obtain
$Q\approx\rm{4}$ ($I_c=\rm{200\;nA}$ and $C_J=\rm{100\;fF}$),
which yields $I_m\simeq 0.35I_c$ through (\ref{eq:Im}) like in the
experiment. In the diagram of Fig.~\ref{fig:phasediagram} we also
present $I_c$ of the dc-SQUID at fluxes 0, $ \pm 0.28\Phi _0$ and
$ \pm 0.41\Phi _0$ by horizontal dashed lines. It can be seen that
the intersections of the dashed lines and the boundary of the UPD
regime are very close to the experimental values of $T_D$. The
saturation of the histograms and their re-entrant steepness  is
thus a manifestation of the cross-over from TA escape into UPD due
to dissipation. In the case of a single JJ with $R(\omega
_p) \simeq\rm{680\;\Omega}$ we obtain $Q\approx\rm{13}$ yielding $I_m\approx{\rm 0.1\;}I_c$ and $T_D \approx 1.2$ K,
instead of the measured ${\rm 0.3\;}I_c$ and 650 mK. By taking
$R(\omega _p)\simeq\rm{230\;\Omega}$ we obtain $Q = 4.4$, yielding
$I_m\approx{\rm 0.3\;}I_c$ and $T_D \approx 700$ mK instead, which
is consistent with what we obtain from the position $I_m/I_c$.

Fig.~\ref{fig:histot} shows that the plain TA-MQT models cannot
account for our observations. Except for the data at the lowest
temperatures the width and the position of the measured histograms
deviate from simulated ones (dotted line). Dissipation alone
cannot explain the difference. The basic TA model yields $\Delta I
\propto T^{2/3}$ \cite{Weiss} and it can be seen in
Figs.~\ref{fig:histot} and \ref{fig:I50&dI} that the dc-SQUID has
weaker temperature dependence even well below $T_D$. In the dc-SQUID
there are just few energy levels in the well and thus the
assumptions of continuous energy spectrum are not valid here
\cite{Weiss}. The semiclassical model of Larkin and Ovchinnikov
\cite{Larkin86} takes into account the influence of the shape of
the potential, in particular the fact that it is not harmonic. The
total escape probability is calculated using
$P_{esc}(\tau)=1-\sum_k \rho _k(\tau)$, where $\rho _k(\tau)$ is
the probability of finding the particle in a state $k$ after the
current pulse of length $\tau$. The kinetic equation of the phase
particle can be written as $\frac{d\rho _k}{dt}= \sum_j(\gamma
_{kj}\rho _j-\gamma _{jk}\rho _k) -\Gamma_{k} \rho _k$. We take
into account only transitions $\gamma _{jk}$ between  neighboring
levels and tunnelling out, $\Gamma _k$. The relaxation rates
between levels $j$ and $j-1$ are well approximated by $\gamma
_{j-1,j}=j \omega _p /4Q$. We assume detailed balance and write
$\gamma_{j,j-1}$=$ e^{-\beta (E_j-E_{j-1})}\gamma_{j-1,j}$. The
positions and the escape rates are calculated using the results in
Ref.~\cite{Larkin86}. The final state $\overline{\rho} \equiv \left[ \rho _1 \; \rho _2 \; \ldots \; \rho _k
\right] $ is calculated by numerically using
 $\overline{\rho}(\tau)= \frac{1}{\tau} \int_{0}^{\tau}e^{\mathbf{A}(t)}\overline{\rho}(0) dt$, where
 $\mathbf{A}$ is the transition matrix.
$I$ is set to zero in the beginning, and 
$\overline{\rho}(0)$ is Boltzmann distributed.

The effect of the relocalizing dissipation can be taken into account also in the
quantized energy level model. Using again $E_D=8 E_J/Q$, and the fact that the energy
difference between the two successive maxima is $- 2 \pi E_J I/I_c $,
 the level energy $E$ must satisfy
\begin{equation}
\label{eq:Ed_cond} \Delta U-E<E_J \left( 2\pi I/I_c-8/Q \right)
\end{equation}
to allow switching into the free running state. If
(\ref{eq:Ed_cond}) is violated, the corresponding tunnelling rate
is set to zero \cite{footnote}.   Note that (\ref{eq:Ed_cond})
gives the same threshold as (\ref{eq:Im}) for  $\Delta U= E$, but
in Eq.~(\ref{eq:Ed_cond})  we assume that after tunnelling the
starting point is not at the potential maximum. The solid red
lines in Fig. \ref{fig:histot} present results of simulations with
quantized energy levels and dissipation for the data of the
dc-SQUID at zero and $\pm 0.28\Phi _0$ fluxes. At $\pm 0.41\Phi
_0$  $I_c$ is so small that the escape probability is large even
at zero bias (except at the lowest temperatures). This means that
the phase is moving constantly rather than infrequently escaping
from a localized state, and thus our  model  does not work
anymore. The only fitting parameter  was $Q$, and the fitted
values were in a very reasonable range. At $\Phi=0$ we find
$Q\approx 6$ at the lowest temperature, and it decreases with
increasing temperature up to 4 at $\rm{325\;mK}$. At $\pm 0.28\Phi
_0$ $Q\approx 3$ to 4, again decreasing with temperature. In Fig.
\ref{fig:I50&dI} is presented the $I_{50\%}$ and $\Delta I$
parameters of the simulated histograms. We present also the
simulated parameters for single junction ($Q\approx 4$)
\cite{note2}. In the measurements on a single junction, the number
of energy levels is large ($\approx \rm{20}$) and the level
separation is smaller than the level width. Our model assumes
well-separated levels and thus the agreement with a simple TA
model is better for a single junction at temperatures $T < T_D$,
especially in Fig. \ref{fig:I50&dI} (a),
whereas re-entrant steepness of the histograms could not be
explained with the basic TA model. The agreement between the
simulation and the measurements on a DC-SQUID is excellent. The
position and the width of the measured histograms coincide and, in
particular, at higher temperatures the simulated histograms for
both samples also start to get steeper again. At higher
temperatures the upper energy levels, whose escape rate is
significant with smaller potential tilting angles, are populated
as well. The histograms thus peak at smaller currents and the
condition (\ref{eq:Ed_cond}) is not necessarily fulfilled anymore.
What remains in the measured (and simulated) histograms is the
escape from the states at the tail of the Boltzmann distribution
above the dissipation barrier.

The Academy of Finland, EU IST-FET-SQUBIT2 and the French ACI,
IPMC and IUF programs are acknowledged for financial support. We
are grateful to A. O. Niskanen,  T.T. Heikkil\"a and M.A. Paalanen
for helpful discussions.


\begin{thebibliography}{99}
\bibitem{footnote} With the low $Q$ values in the experiment, the phase relaxes in the next well in
                   a time $\sim \omega _p^{-1}\sim\rm{100\; ps}$, which is far shorter than the typical
                   time interval between phase diffusion events with studied current bias values ($\rm{10...100\;\mu s}$).
\bibitem{note2} We have not taken into account in the analysis that the superconducting gap is reduced when $T$ is
                approaching $T_c \approx\rm{1.2\;K}$.
\end{thebibliography}
\end{document}